# First-Principles-Based Grand Unified Theory (GUT) for Micro–Macro Modal Quantization (MQ)—Part V: Planck–Einstein–de Broglie Eigen-Relations, Act 1

Ren-Zun Lian*
*School of Electronic Engineering*
*Xidian University*
Xi'an, China
rzlian@vip.163.com

***Abstract*—As a continuation of our previous papers, the Planck–Einstein eigen-relation (or alternatively called Lagrange–Hamilton eigen-relation) and de Broglie eigen-relation in macroscopic electrodynamics are revealed in this paper. Using the macroscopic Planck–Einstein eigen-relation, the generalized quality factor defined in our previous papers can be derived naturally and rigorously. These above provide a solid physical and mathematical foundation for defining the concept of generalized quality factor, whose physical meaning is the reciprocal of normalized free-photon number. Using the above these, we also derive a novel and effective calculation formulation for the generalized quality factor. Based on the macroscopic Planck–Einstein eigen-relation and generalized quality factor, a series of conclusions regarding macroscopic modal quantization (MMQ) are also derived, for example: the physical meaning of eigen-value is further clarified from several different points of view; a natural normalization method for eigenmodal quanta is obtained; a generalized Parseval identity is formulated; the lower bound of generalized quality factor is revealed.**

## I. Introduction

In 1900, Planck [1], [2] proposed the famous quantum hypothesis. In 1905, Einstein [3] established the following famous **Planck–Einstein relation**:

$$E = \hbar\omega , \tag{1}$$

to explain photoelectric effect. Historically, Planck–Einstein relation (1) directly correlates the frequency of light with the discrete energy of individual photons, and then first elucidates the wave-particle duality of light. The Planck–Einstein relation (1) invalidated the classical postulate of continuous energy, and then heralded the pivotal transition from classical macroscopic physics to modern quantum physics.

Obviously, the Planck–Einstein relation (1) establishes a relationship between the microscopic energy $E$ and the microscopic action $\hbar$ (i.e., reduced Planck constant). Recently, in macroscopic electrodynamics we also discovered a similar relation [4]:

$$\boxed{\tilde{\mathcal{H}}_0 = \frac{1}{\theta} \cdot \mathcal{L}_0} , \tag{2}$$

which establishes the relationship between the macroscopic energy—time-independent generalized Hamiltonian $\tilde{\mathcal{H}}_0$—and the macroscopic action (multiplying a coefficient $\omega$)—time-independent/time-averaged Lagrangian $\mathcal{L}_0$. For the convenience of the following discussions, we call the (2) **Lagrange–Hamilton (or action–energy) eigen-relation**.

By transforming the generalized Hamiltonian $\tilde{\mathcal{H}}_0$ and the classical Lagrangian $\mathcal{L}_0$ in (2) into the corresponding operators $\tilde{\mathrm{H}}_0$ and $\mathrm{L}_0$, [4] gave the following **Lagrange–Hamilton eigen-equation**:

$$\tilde{\mathrm{H}}_0 \left| \psi_\xi^{(0)} \right\rangle = \frac{1}{\theta_\xi^{(0)}} \cdot \mathrm{L}_0 \left| \psi_\xi^{(0)} \right\rangle , \tag{3}$$

and used the above eigen-equation to calculate the macroscopic eigenmodal quanta (i.e., eigenmodes) in macroscopic electrodynamic problem, and then achieved the macroscopic modal quantization (MMQ) for macroscopic electrodynamics.

In the Parts II~IV of [4], it was revealed that: the Lagrange–Hamilton eigen-equation is equivalent to the macroscopic versions of Heisenberg equation, Schrödinger equation, and Feynman path integral formulation. Thus, the Lagrange–Hamilton eigen-relation (2) and eigen-equation (3) have become the **first principles** in electrodynamic MMQ.

Based on the above observations, we reasonably infer that: the Lagrange–Hamilton eigen-relation (2) is just the macroscopic counterpart of the Planck–Einstein relation (1). The central purpose of this paper is to discuss the micro–macro correspondence relationship between (1) and (2), and uses the Lagrange–Hamilton eigen-relation (or called **macroscopic Planck–Einstein eigen-relation**) to formulate the generalized quality factor $\Theta$ in macroscopic electrodynamics. Focusing on this purpose, we organize this paper as follows:

- Sections II and III reveal the equivalence relationships between microscopic Planck–Einstein–de Broglie relations and macroscopic action–energy–momentum eigen-relations;

This work was conducted without support from any funding agency or grant, and all related expenses are personally borne by the author.
A preprint version of this paper was publicly posted on arXiv on June 2026 under the "arXiv.org perpetual, non-exclusive license" (arXiv: 1804.09246v4, available at https://arxiv.org/abs/1804.09246v4).

- ◊ Sections II-A and II-B derive the macroscopic action–energy eigen-relation (2) (i.e., Lagrange–Hamilton eigen-relation) and macroscopic action–momentum eigen-relation (i.e., the scalar version of [5, Sec. IV]) from the microscopic Planck–Einstein relation (1) and microscopic de Broglie relation [6]–[8], respectively;
- ◊ Section II-C casts the eigen-relations into the corresponding eigen-equations;
- ◊ Section III derives the eigen-relations from the eigen-equations;
- ◊ Section III also derives the microscopic Planck–Einstein relation (1) and de Broglie relation from the macroscopic action–energy eigen-relation (2) and action–momentum eigen-relation, respectively;

- Section IV revisits the generalized quality factor Θ defined in [9, Eq. (14)] and [10, Eq. (18)] from the perspective of Lagrange–Hamilton eigen-relation (2);
- Section V gives some further discussions related to MMQ and generalized quality factor Θ;
- Section VI provides some typical numerical experiments;
- Section VII concludes and summarizes the results obtained in this paper.

In what follows, the time factor $e^{-i\omega t}$ and the inner-product form $<\boldsymbol{f}\,|\,\boldsymbol{g}>_{\text{L/S/V}}=\int_{\text{L/S/V}} \mathrm{d}^{1/2/3}x \boldsymbol{f}^{*}\cdot\boldsymbol{g}$ are used throughout, where the superscript " $*$ " is the conjugate operation. The transpose conjugate operation for matrix and vector is denoted by a superscript dagger " $\dagger$ ".

## II. From Microscopic Planck–Einstein–de Broglie Relations to Macroscopic Action–Energy–Momentum Eigen-Relations and Eigen-Equations

In this section, we focus on revealing the Planck–Einstein eigen-relation in macroscopic electrodynamics (or called macroscopic Planck–Einstein or Lagrange–Hamilton or action–energy eigen-relation) and the de Broglie eigen-relation in macroscopic electrodynamics (or called macroscopic de Broglie or action–momentum eigen-relation, which is the scalar version of the one given in [5, Sec. IV]).

### *A. From Microscopic Planck–Einstein Relation to Macroscopic Action–Energy Eigen-Relation*

Inspired by our previous papers [4], we directly replace the $E$ and $\hbar$ in (1) with $\tilde{\mathcal{H}}_0$ and $\bar{\mathcal{S}}_0/\theta$, and then we have that

$$
\begin{array}{cccccc}
\tilde{\mathcal{H}}_0 & & & & & \\
\Updownarrow & & & & & \\
E & = & \hbar & \cdot & \omega & . \\
 & & \updownarrow & & & \\
 & & \theta^{-1}\cdot\bar{\mathcal{S}}_0 & & &
\end{array}
\tag{4}
$$

i.e.,

$$
\boxed{\omega\cdot\bar{\mathcal{S}}_0=\theta\cdot\tilde{\mathcal{H}}_0}\,,
\tag{5}
$$

where the $\bar{\mathcal{S}}_0$ is the time-independent/zero-order action (reduced version). The above (5) is just the so-called **macroscopic Planck–Einstein eigen-relation**, or called **macroscopic action–energy eigen-relation**.

Employing the relationship $\bar{\mathcal{S}}_0=\mathcal{S}_0/(2\pi)=\mathcal{L}_0/\omega$ [4], the (5) can be transformed as follows:

$$
\tilde{\mathcal{H}}_0=\frac{1}{\theta}\cdot\frac{\mathcal{L}_0}{\omega}\cdot\omega=\frac{1}{\theta}\cdot\mathcal{L}_0\,,
\tag{6}
$$

i.e., (2). Thus,

$$
\boxed{\mathcal{L}_0=\theta\cdot\tilde{\mathcal{H}}_0}\,,
\tag{7}
$$

which is just the so-called **macroscopic Lagrange–Hamilton eigen-relation**, which is completely equivalent to (5).

From the macroscopic Lagrange–Hamilton eigen-relation (7), we also have the following relationship:

$$
\underbrace{\left(\tau/2\right)^{-1}\cdot\mathcal{L}_0}_{\mathcal{P}_0^-}=\theta\cdot\underbrace{\left(\tau/2\right)^{-1}\cdot\tilde{\mathcal{H}}_0}_{\mathcal{P}_0^+}\,,
\tag{8}
$$

i.e.,

$$
\boxed{\mathcal{P}_0^-=\theta\cdot\mathcal{P}_0^+}\,,
\tag{9}
$$

which can be called $\boldsymbol{\mathcal{P}_0^+-\mathcal{P}_0^-}$ **eigen-relation**. Here, the relationships $(\tau/2)^{-1}\cdot\tilde{\mathcal{H}}_0=\mathcal{P}_0^+$ and $(\tau/2)^{-1}\cdot\mathcal{L}_0=\mathcal{P}_0^-$ given in [4] are used, where the $\tau$ is the time period of field.

### *B. From Microscopic de Broglie Relation to Macroscopic Action–Momentum Eigen-Relation*

As every physicist knows, the Planck–Einstein relation (1) and the following **Einstein mass–energy relation** [11]

$$
E=mc^2=pc
\tag{10}
$$

together give the following **de Broglie relation** [6]–[8]

$$
p=\hbar\frac{\omega}{c}=\hbar k\,,
\tag{11}
$$

which is well-known in quantum physics.

Similarly to (4), we have the following micro–macro correspondence relationship:

$$
\begin{array}{cccccc}
 & & \mathcal{G}_0 & & & \\
 & & \Updownarrow & & & \\
c & \cdot & p & = & \hbar & \cdot & ck \\
 & & & = & \hbar & \cdot & \omega \;\;, \\
 & & & & \updownarrow & & \\
 & & & & \theta^{-1}\cdot\bar{\mathcal{S}}_0 & &
\end{array}
\tag{12}
$$

where we have used the relation $ck=c(2\pi/\lambda)=c(2\pi\nu/c)=\omega$. The $c$ is the velocity of light, and

$$
\mathcal{G}_0=\frac{1}{c}\cdot\tilde{\mathcal{H}}_0\,,
\tag{13}
$$

where the $\mathcal{G}_0$ is so-called time-independent momentum carried by macroscopic electromagnetic field.

From (5) and (13), we immediately have the following **macroscopic de Broglie eigen-relation**:

$$\boxed{\omega \cdot \bar{\mathcal{S}}_0 = \theta \cdot c \cdot \mathcal{G}_0}\,, \tag{14}$$

or called **macroscopic action–momentum eigen-relation**. From (7) and (13), we immediately have the following

$$\boxed{c^{-1} \cdot \mathcal{L}_0 = \theta \cdot \mathcal{G}_0}\,, \tag{15}$$

which can be viewed as an alternative eigen-relation between macroscopic action (multiplying a coefficient $\omega$ ) and macroscopic momentum.

### *C. From Eigen-Relations to Eigen-Equations*

Inspired by the convention of quantum physics, the transformation from eigen-relation to eigen-equation is given in this section directly.

By casting the above eigen-relations (5)/(7), (9), and (14)/(15) into their operator forms, we immediately obtain the following eigen-equations [4], [5], [9], [10]:

$$\left.\begin{aligned} \mathrm{L}_0 \left|\psi_\xi^{(0)}\right\rangle &= \theta_\xi^{(0)} \cdot \tilde{\mathrm{H}}_0 \left|\psi_\xi^{(0)}\right\rangle, \\ \mathrm{P}_0^- \left|\psi_\xi^{(0)}\right\rangle &= \theta_\xi^{(0)} \cdot \mathrm{P}_0^+ \left|\psi_\xi^{(0)}\right\rangle, \\ c^{-1} \cdot \mathrm{L}_0 \left|\psi_\xi^{(0)}\right\rangle &= \theta_\xi^{(0)} \cdot \mathrm{G}_0 \left|\psi_\xi^{(0)}\right\rangle. \end{aligned}\right\} \tag{16}$$

Here, the first eigen-equation [4] is completely equivalent to (3); the second eigen-equation is just the classical eigen-equation [5], [9], [10] used to do MMQ; the third eigen-equation is a novel form. In fact, the above three eigen-equations are equivalent to each other, because

$$\left.\begin{aligned} \tilde{\mathrm{H}}_0 &= (\tau/2) \cdot \mathrm{P}_0^+, \\ \mathrm{L}_0 &= (\tau/2) \cdot \mathrm{P}_0^-, \\ \mathrm{G}_0 &= c^{-1} \cdot \tilde{\mathrm{H}}_0, \end{aligned}\right\} \tag{17}$$

where the first two relationships were derived in [4].

Here, the superscript " (0) " used in eigenmodal quanta $|\psi_\xi^{(0)}>$ and eigen-values $\theta_\xi^{(0)}$ had been explained in the Part III of [4]: the $|\psi_\xi^{(0)}>$ is the stationary-state eigenmodal quanta, and the $\theta_\xi^{(0)}$ is the corresponding stationary-state eigen-values. For simplifying the symbolic system of the following parts of this paper, the superscript " (0) " will be omitted.

## III. From Macroscopic Action–Energy–Momentum Eigen-Relations and Eigen-Equations to Microscopic Planck–Einstein–de Broglie Relations

In the above Section II, we have exhibited the following important facts:

1) the microscopic Planck–Einstein relation (1) has its macroscopic version—the action–energy eigen-relation (5), or equivalently written as the Lagrange–Hamilton eigen-relation (7), which is equivalent to the classical $\mathcal{P}_0^+ - \mathcal{P}_0^-$ eigen-relation (9);
2) the microscopic de Broglie relation (11) has its macroscopic version—the action–momentum eigen-relation (14), or equivalently written as the action–momentum eigen-relation (15);
3) the Lagrange–Hamilton eigen-relation (7) and action–momentum eigen-relation (15) directly give the Lagrange–Hamilton eigen-equation [the first line in (16)] and action–momentum eigen-equation [the third line in (16)], respectively;
4) the Lagrange–Hamilton and action–momentum eigen-equations are equivalent to the classical $\mathcal{P}_0^+ - \mathcal{P}_0^-$ eigen-equation [the second line in (16)].

In summary,

$$\begin{Bmatrix} (1) \\ (9) \\ (11) \end{Bmatrix} \xRightarrow[p \Leftrightarrow \mathcal{G}_0,\ \hbar \leftrightarrow \bar{\mathcal{S}}_0/\theta]{E \Leftrightarrow \tilde{\mathcal{H}}_0,\ \hbar \leftrightarrow \bar{\mathcal{S}}_0/\theta} \begin{Bmatrix} (5) \\ (9) \\ (14) \end{Bmatrix} \xLeftrightarrow[\bar{\mathcal{S}}_0 = \mathcal{L}_0/\omega]{\bar{\mathcal{S}}_0 = \mathcal{L}_0/\omega} \begin{Bmatrix} (7) \\ (9) \\ (15) \end{Bmatrix} \xRightarrow{\text{operatorization}} (16)\,. \tag{18}$$

In this section, we will prove that the Planck–Einstein relation (1), $\mathcal{P}_0^+ - \mathcal{P}_0^-$ eigen-relation (9), and de Broglie relation (11) can also be derived from the eigen-equations given in (16).

The converse statement of the above (18) is presented below.

$$\begin{aligned} &\begin{matrix}\text{eigen-equations} \\ (16)\end{matrix} \xRightarrow{(17)} \left\{ \begin{matrix} \overbrace{\langle\psi|\mathrm{L}_0|\psi\rangle}^{\mathcal{L}_0} = \theta \cdot \overbrace{\langle\psi|\tilde{\mathrm{H}}_0|\psi\rangle}^{\tilde{\mathcal{H}}_0} \\ \Updownarrow \\ \underbrace{\langle\psi|\mathrm{P}_0^-|\psi\rangle}_{\mathcal{P}_0^-} = \theta \cdot \overbrace{\langle\psi|\mathrm{P}_0^+|\psi\rangle}^{\mathcal{P}_0^+} \\ \Updownarrow \\ c^{-1} \cdot \underbrace{\langle\psi|\mathrm{L}_0|\psi\rangle}_{\mathcal{L}_0} = \theta \cdot \underbrace{\langle\psi|\mathrm{G}_0|\psi\rangle}_{\mathcal{G}_0} \end{matrix} \right\} \\ &\Leftrightarrow \left\{ \begin{matrix} \overbrace{\text{Lagrange} - \text{Hamilton}}^{\mathcal{L}_0 - \tilde{\mathcal{H}}_0} \text{ eigen-relation (7)} \\ \Updownarrow \\ \mathcal{P}_0^+ - \mathcal{P}_0^- \text{ eigen-relation (9)} \\ \Updownarrow \\ \underbrace{\text{action} - \text{momentum}}_{\mathcal{L}_0 - \mathcal{G}_0} \text{ eigen-relation (15)} \end{matrix} \right\} \\ &\xLeftrightarrow[\bar{\mathcal{S}}_0 = \frac{\mathcal{L}_0}{\omega}]{\bar{\mathcal{S}}_0 = \frac{\mathcal{L}_0}{\omega}} \left\{ \begin{matrix} \overbrace{\text{action} - \text{energy}}^{\bar{\mathcal{S}}_0 - \tilde{\mathcal{H}}_0} \text{ eigen-relation (5)} \\ \Updownarrow \\ \mathcal{P}_0^+ - \mathcal{P}_0^- \text{ eigen-relation (9)} \\ \Updownarrow \\ \underbrace{\text{action-momentum}}_{\bar{\mathcal{S}}_0 - \mathcal{G}_0} \text{ eigen-relation (14)} \end{matrix} \right\} \\ &\xRightarrow[p \Leftrightarrow \mathcal{G}_0,\ \hbar \leftrightarrow \bar{\mathcal{S}}_0/\theta]{E \Leftrightarrow \tilde{\mathcal{H}}_0,\ \hbar \leftrightarrow \bar{\mathcal{S}}_0/\theta} \left\{ \begin{matrix} \text{Planck} - \text{Einstein relation (1)} \\ \Updownarrow \\ \mathcal{P}_0^+ - \mathcal{P}_0^- \text{ eigen-relation (9)} \\ \Updownarrow \\ \text{de Broglie relation (11)} \end{matrix} \right\} \\ &\Leftrightarrow \text{Einstein mass} - \text{energy relation (10).} \end{aligned} \tag{19}$$

Here, the last "$\Leftrightarrow$" is based on $E = \hbar\omega = \hbar kc = pc = mc^2$; the other "$\Rightarrow$" and "$\Leftrightarrow$" are obvious.

Up to this point, we have obtained the following set of important relationships:

$$\boxed{(10) \overset{(11)}{\underset{(1)}{\Leftrightarrow}} \left\{ \begin{array}{c} (1) \\ \Updownarrow \\ (9) \\ \Updownarrow \\ (11) \end{array} \right\} \Leftrightarrow \left\{ \begin{array}{c} (5) \\ \Updownarrow \\ (9) \\ \Updownarrow \\ (14) \end{array} \right\} \Leftrightarrow \left\{ \begin{array}{c} (7) \\ \Updownarrow \\ (9) \\ \Updownarrow \\ (15) \end{array} \right\} \Leftrightarrow (16)}, \quad (20)$$

where the first "$\Leftrightarrow$" means the well-known relationships $(10)\,\&\,(11) \Rightarrow (1)$ and $(1)\,\&\,(10) \Rightarrow (11)$ and $(1)\,\&\,(11) \Rightarrow (10)$.

## IV. Generalized Quality Factor Θ Revisited

When an open electromagnetic system works at time-harmonic state (or called stationary state), its quality factor Q is very difficult to calculate, and many scholars [12]–[24] devoted strenuous efforts to this problem, but the problem has not yet been completely solved. In [9] and [10], the authors proposed an alternative perspective to define the quality factor, and introduced a concept of **generalized quality factor Θ**.

### A. Generalized Quality Factor Θ in [9] and [10]

The generalized quality factor Θ proposed in [9, Eq. (14)] and [10, Eq. (18)] is defined and calculated by using the so-called modal three-component eigen-decomposition as follows. First, any stationary state $|\psi>$ is decomposed into three components—purely capacitive component $|\psi_{\text{cap}}>$, purely resonant component $|\psi_{\text{res}}>$, and purely inductive component $|\psi_{\text{ind}}>$—as follows [25, Eq. (22)]:

$$\left|\psi\right\rangle = \left|\psi_{\text{cap}}\right\rangle + \left|\psi_{\text{res}}\right\rangle + \left|\psi_{\text{ind}}\right\rangle . \quad (21)$$

Here, the $|\psi_{\text{cap}}>$ is the linear superposition of all of the capacitive eigenmodal quanta; the $|\psi_{\text{res}}>$ is the linear superposition of all of the resonant eigenmodal quanta; the $|\psi_{\text{ind}}>$ is the linear superposition of all of the inductive eigenmodal quanta. Using the three-component eigen-decomposition, [9, Eq. (14)] and [10, Eq. (18)] defined and calculated the generalized quality factor Θ as follows:

$$\begin{aligned} \Theta &= \frac{\overbrace{\left\langle \psi_{\text{cap}} \middle| \mathbb{P}_0^- \middle| \psi_{\text{cap}} \right\rangle}^{>0} + \overbrace{\left\langle \psi_{\text{res}} \middle| \mathbb{P}_0^- \middle| \psi_{\text{res}} \right\rangle}^{=0} - \overbrace{\left\langle \psi_{\text{ind}} \middle| \mathbb{P}_0^- \middle| \psi_{\text{ind}} \right\rangle}^{<0}}{\underbrace{\left\langle \psi_{\text{cap}} + \psi_{\text{res}} + \psi_{\text{ind}} \middle| \mathbb{P}_0^+ \middle| \psi_{\text{cap}} + \psi_{\text{res}} + \psi_{\text{ind}} \right\rangle}_{>0}} \\ &= \frac{\left\langle \psi_{\text{cap}} \middle| \mathbb{P}_0^- \middle| \psi_{\text{cap}} \right\rangle - \left\langle \psi_{\text{ind}} \middle| \mathbb{P}_0^- \middle| \psi_{\text{ind}} \right\rangle}{\left\langle \psi \middle| \mathbb{P}_0^+ \middle| \psi \right\rangle}. \end{aligned} \quad (22)$$

The building blocks $<\psi|\mathbb{P}_0^+|\psi>$, $<\psi_{\text{cap}}|\mathbb{P}_0^-|\psi_{\text{cap}}>$, and $<\psi_{\text{ind}}|\mathbb{P}_0^-|\psi_{\text{ind}}>$ used to calculate Θ are easy to be obtained, if the MMQ for open electromagnetic system has been finished, and the three-component eigen-decomposition has been done.

The **difference** between the previous [9, Eq. (14)] and [10, Eq. (18)] and the above (22) is due to that: literatures [9] and [10] used the time factor $e^{j\omega t}$, while this series of papers ([4] and this paper) use the time factor $e^{-i\omega t}$. In the following parts of this section, we focus on revealing the physical and mathematical foundations for defining the generalized quality factor Θ as (22).

### B. Preliminaries for Sections IV-C and IV-D

If the $\tilde{\mathbb{H}}_0$ is positively definite at frequency $\omega$, there must be a non-singular matrix $\mathbb{T}$, such that $\mathbb{T}^\dagger \cdot \tilde{\mathbb{H}}_0 \cdot \mathbb{T}$ becomes a scalar matrix $\mathbb{I} \cdot \hbar\omega$, i.e., [4]

$$\left. \begin{aligned} \mathbb{T}^\dagger \cdot \tilde{\mathbb{H}}_0 \cdot \mathbb{T} &= \tilde{\mathbb{H}}_0^{\mathrm{T}} \\ &= \mathbb{I} \cdot \hbar\omega, \\ \mathbb{T}^\dagger \cdot \mathbb{L}_0 \cdot \mathbb{T} &= \mathbb{L}_0^{\mathrm{T}}, \\ \mathbb{T}^{-1}\left|\psi\right\rangle &= \left|\psi^{\mathrm{T}}\right\rangle, \end{aligned} \right\} \quad (23)$$

and then the Lagrange–Hamilton eigen-equation, i.e., the first one in (16), becomes the following form:

$$\begin{aligned} \overbrace{\mathbb{T}^\dagger \cdot \mathbb{L}_0 \cdot \mathbb{T}}^{\mathbb{L}_0^{\mathrm{T}}} \cdot \overbrace{\mathbb{T}^{-1}\left|\psi_\xi\right\rangle}^{\left|\psi_\xi^{\mathrm{T}}\right\rangle} &= \theta_\xi \cdot \overbrace{\mathbb{T}^\dagger \cdot \tilde{\mathbb{H}}_0 \cdot \mathbb{T}}^{\mathbb{I}\cdot\hbar\omega} \cdot \overbrace{\mathbb{T}^{-1}\left|\psi_\xi\right\rangle}^{\left|\psi_\xi^{\mathrm{T}}\right\rangle} \\ &= \hbar\omega \cdot \theta_\xi \cdot \left|\psi_\xi^{\mathrm{T}}\right\rangle, \end{aligned} \quad (24)$$

where the superscript "T" means that the corresponding quantities are derived from using the $\mathbb{T}$-based transformation.

Based on (24), there must be an unitary matrix $\mathbb{U}$, such that the $\mathbb{U}^\dagger \cdot \mathbb{L}_0^{\mathrm{T}} \cdot \mathbb{U}$ becomes a diagonal matrix $\Lambda^{\mathrm{TU}}$, i.e.,

$$\underbrace{\mathbb{U}^\dagger \cdot \mathbb{L}_0^{\mathrm{T}} \cdot \mathbb{U}}_{\mathbb{L}_0^{\mathrm{TU}}} = \Lambda^{\mathrm{TU}} = \hbar\omega \cdot \operatorname{diag}\left\{\theta_1, \theta_2, \cdots\right\} . \quad (25)$$

By defining (if $\theta_1, \theta_2, \cdots \neq 0$)

$$\mathbb{V} \overset{\text{def}}{=\!=} \mathbb{U} \cdot \left(\hbar\omega\right)^{1/2} \left|\Lambda^{\mathrm{TU}}\right|^{-1/2} = \mathbb{U} \cdot \operatorname{diag}\left\{|\,\theta_1\,|^{-1/2}, |\,\theta_2\,|^{-1/2}, \cdots\right\}, \quad (26)$$

it is easy to have

$$\underbrace{\mathbb{V}^\dagger \cdot \mathbb{L}_0^{\mathrm{T}} \cdot \mathbb{V}}_{\mathbb{L}_0^{\mathrm{TV}}} = \Lambda_{\pm\hbar\omega}^{\mathrm{TV}} = \hbar\omega \cdot \operatorname{diag}\left\{\operatorname{sign}\theta_1, \operatorname{sign}\theta_2, \cdots\right\}, \quad (27)$$

where the $\Lambda_{\pm\hbar\omega}^{\mathrm{TV}}$ is similar to the $\Lambda_{\pm\hbar\omega}^{\Gamma}$ in [4, Part II, Eq. (52)].

Similarly to the above $\mathbb{L}_0^{\mathrm{T}}$-oriented transformations, we have the following transformations:

$$\left. \begin{aligned} \tilde{\mathbb{H}}_0^{\mathrm{TU}} &\overset{\text{def}}{=\!=} \mathbb{U}^\dagger \cdot \tilde{\mathbb{H}}_0^{\mathrm{T}} \cdot \mathbb{U}, \\ &=\!= \mathbb{I} \cdot \hbar\omega \\ \left|\psi^{\mathrm{TU}}\right\rangle &\overset{\text{def}}{=\!=} \mathbb{U}^{-1}\left|\psi^{\mathrm{T}}\right\rangle, \end{aligned} \right\} \quad (28)$$

and

$$\left. \begin{aligned} \tilde{\mathbb{H}}_0^{\mathrm{TV}} &\overset{\text{def}}{=\!=} \mathbb{V}^\dagger \cdot \tilde{\mathbb{H}}_0^{\mathrm{T}} \cdot \mathbb{V}, \\ &=\!= \left(\hbar\omega\right)^2 \left|\Lambda^{\mathrm{TU}}\right|^{-1} \\ \left|\psi^{\mathrm{TV}}\right\rangle &\overset{\text{def}}{=\!=} \mathbb{V}^{-1}\left|\psi^{\mathrm{T}}\right\rangle. \end{aligned} \right\} \quad (29)$$

The $\mathbb{V}$ here needs not be a unitary transformation.

For any two modes $|\psi_\xi>$ and $|\psi_\zeta>$ (which need not be eigenmodal quanta), there exist the following relationships:

$$\left.\begin{aligned}
\left\langle \psi_\xi \middle| \tilde{\mathrm{H}}_0 \middle| \psi_\zeta \right\rangle &= \underbrace{\left\langle \psi_\xi \right| \left(\mathrm{T}^\dagger\right)^{-1}}_{\left\langle \psi_\xi^{\mathrm{T}} \right|} \cdot \underbrace{\overbrace{\mathrm{T}^\dagger \cdot \tilde{\mathrm{H}}_0 \cdot \mathrm{T}}^{\mathrm{I}\cdot\hbar\omega}}_{\tilde{\mathrm{H}}_0^{\mathrm{T}}} \cdot \underbrace{\mathrm{T}^{-1}\left|\psi_\zeta\right\rangle}_{\left|\psi_\zeta^{\mathrm{T}}\right\rangle} \\
&= \underbrace{\left\langle \psi_\xi^{\mathrm{T}} \right| \left(\mathrm{U}^\dagger\right)^{-1}}_{\left\langle \psi_\xi^{\mathrm{TU}} \right|} \cdot \underbrace{\overbrace{\mathrm{U}^\dagger \cdot \tilde{\mathrm{H}}_0^{\mathrm{T}} \cdot \mathrm{U}}^{\mathrm{I}\cdot\hbar\omega}}_{\tilde{\mathrm{H}}_0^{\mathrm{TU}}} \cdot \underbrace{\mathrm{U}^{-1}\left|\psi_\zeta^{\mathrm{T}}\right\rangle}_{\left|\psi_\zeta^{\mathrm{TU}}\right\rangle} \\
&= \underbrace{\left\langle \psi_\xi^{\mathrm{T}} \right| \left(\mathrm{V}^\dagger\right)^{-1}}_{\left\langle \psi_\xi^{\mathrm{TV}} \right|} \cdot \underbrace{\overbrace{\mathrm{V}^\dagger \cdot \tilde{\mathrm{H}}_0^{\mathrm{T}} \cdot \mathrm{V}}^{(\hbar\omega)^2\left|\Lambda^{\mathrm{TU}}\right|^{-1}}}_{\tilde{\mathrm{H}}_0^{\mathrm{TV}}} \cdot \underbrace{\mathrm{V}^{-1}\left|\psi_\zeta^{\mathrm{T}}\right\rangle}_{\left|\psi_\zeta^{\mathrm{TV}}\right\rangle}, \\
\left\langle \psi_\xi \middle| \mathrm{L}_0 \middle| \psi_\zeta \right\rangle &= \overbrace{\left\langle \psi_\xi \right| \left(\mathrm{T}^\dagger\right)^{-1}}^{\left\langle \psi_\xi^{\mathrm{T}} \right|} \cdot \overbrace{\mathrm{T}^\dagger \cdot \mathrm{L}_0 \cdot \mathrm{T}}^{\mathrm{L}_0^{\mathrm{T}}} \cdot \overbrace{\mathrm{T}^{-1}\left|\psi_\zeta\right\rangle}^{\left|\psi_\zeta^{\mathrm{T}}\right\rangle} \\
&= \overbrace{\left\langle \psi_\xi^{\mathrm{T}} \right| \left(\mathrm{U}^\dagger\right)^{-1}}^{\left\langle \psi_\xi^{\mathrm{TU}} \right|} \cdot \underbrace{\overbrace{\mathrm{U}^\dagger \cdot \mathrm{L}_0^{\mathrm{T}} \cdot \mathrm{U}}^{\mathrm{L}_0^{\mathrm{TU}}}}_{\Lambda^{\mathrm{TU}}} \cdot \overbrace{\mathrm{U}^{-1}\left|\psi_\zeta^{\mathrm{T}}\right\rangle}^{\left|\psi_\zeta^{\mathrm{TU}}\right\rangle} \\
&= \overbrace{\left\langle \psi_\xi^{\mathrm{T}} \right| \left(\mathrm{V}^\dagger\right)^{-1}}^{\left\langle \psi_\xi^{\mathrm{TV}} \right|} \cdot \underbrace{\overbrace{\mathrm{V}^\dagger \cdot \mathrm{L}_0^{\mathrm{T}} \cdot \mathrm{V}}^{\mathrm{L}_0^{\mathrm{TV}}}}_{\Lambda_{\pm\hbar\omega}^{\mathrm{TV}}} \cdot \overbrace{\mathrm{V}^{-1}\left|\psi_\zeta^{\mathrm{T}}\right\rangle}^{\left|\psi_\zeta^{\mathrm{TV}}\right\rangle},
\end{aligned}\right\} \tag{30}$$

because of the relationships (23), (25), (27)–(29), $(\mathrm{T}^\dagger)^{-1} = (\mathrm{T}^{-1})^\dagger$, $(\mathrm{U}^\dagger)^{-1} = (\mathrm{U}^{-1})^\dagger = \mathrm{U}$, and $(\mathrm{V}^\dagger)^{-1} = (\mathrm{V}^{-1})^\dagger$.

Based on the above these, we have the following important relationships:

$$\begin{aligned}
\Lambda_{\pm\hbar\omega}^{\mathrm{TV}} \left|\psi_\xi^{\mathrm{TV}}\right\rangle &= \overbrace{\mathrm{V}^\dagger \cdot \mathrm{L}_0^{\mathrm{T}} \cdot \mathrm{V}}^{\Lambda_{\pm\hbar\omega}^{\mathrm{TV}}} \cdot \underbrace{\mathrm{V}^{-1}\left|\psi_\xi^{\mathrm{T}}\right\rangle}_{\left|\psi_\xi^{\mathrm{TV}}\right\rangle} \\
&= \hbar\omega \cdot \theta_\xi \cdot \mathrm{V}^\dagger \left|\psi_\xi^{\mathrm{T}}\right\rangle \\
&= \hbar\omega \cdot \theta_\xi \cdot \mathrm{V}^\dagger \cdot \mathrm{V} \cdot \underbrace{\mathrm{V}^{-1}\left|\psi_\xi^{\mathrm{T}}\right\rangle}_{\left|\psi_\xi^{\mathrm{TV}}\right\rangle} \\
&= \hbar\omega \cdot \theta_\xi \cdot \left[\left(\hbar\omega\right)^{1/2}\left|\Lambda^{\mathrm{TU}}\right|^{-1/2} \cdot \mathrm{U}^\dagger\right] \\
&\qquad \cdot \left[\mathrm{U} \cdot \left(\hbar\omega\right)^{1/2}\left|\Lambda^{\mathrm{TU}}\right|^{-1/2}\right]\left|\psi_\xi^{\mathrm{TV}}\right\rangle \\
&= \left(\hbar\omega\right)^2 \cdot \theta_\xi \cdot \left|\Lambda^{\mathrm{TU}}\right|^{-1}\left|\psi_\xi^{\mathrm{TV}}\right\rangle.
\end{aligned} \tag{31}$$

Here, the first equality is based on (27) and (29); the second equality is based on (24); the third equality is obvious; the fourth and fifth equalities are based on (26) and $\mathrm{U}^\dagger = \mathrm{U}^{-1}$.

The $\Lambda_{\pm\hbar\omega}^{\mathrm{TV}}$ is a diagonal matrix, and the $|\Lambda^{\mathrm{TU}}|^{-1}$ is a positive diagonal matrix (previously, we have suppose that $\theta_1, \theta_2, \cdots \neq 0$), and the $(\hbar\omega)^2$ is a positive number, so it is immediate to obtain the following **important conclusions**:

1) if $\theta_\xi > 0$, and the $n$th diagonal element of $\Lambda_{\pm\hbar\omega}^{\mathrm{TV}}$ is $-\hbar\omega$, then the $n$th element of $|\psi_\xi^{\mathrm{TV}}>$ must be 0;
2) if $\theta_\xi < 0$, and the $n$th diagonal element of $\Lambda_{\pm\hbar\omega}^{\mathrm{TV}}$ is $+\hbar\omega$, then the $n$th element of $|\psi_\xi^{\mathrm{TV}}>$ must be 0.

The following Section IV-C will use the above conclusions to reveal the physical and mathematical foundation of the generalized quality factor Θ defined in (22).

### *C. Physical and Mathematical Foundation for Θ*

Obviously, (if $\theta_1, \theta_2, \cdots \neq 0$)

$$\frac{\left\langle \psi^{\mathrm{TV}} \middle\| \mathrm{L}_0^{\mathrm{TV}} \middle\| \psi^{\mathrm{TV}} \right\rangle}{\left\langle \psi^{\mathrm{TV}} \middle| \psi^{\mathrm{TV}} \right\rangle} = \frac{\left\langle \psi^{\mathrm{TV}} \middle\| \Lambda_{\pm\hbar\omega}^{\mathrm{TV}} \middle\| \psi^{\mathrm{TV}} \right\rangle}{\left\langle \psi^{\mathrm{TV}} \middle| \psi^{\mathrm{TV}} \right\rangle} = \hbar\omega . \tag{32}$$

The above (32) implies that: the $<\psi^{\mathrm{TV}} | \psi^{\mathrm{TV}}>$ normalizes the $<\psi^{\mathrm{TV}} \| \mathrm{L}_0^{\mathrm{TV}} \| \psi^{\mathrm{TV}}>$ into a single-photon energy $\hbar\omega$.

Based on the above observation, we also use the $<\psi^{\mathrm{TV}} | \psi^{\mathrm{TV}}>$ to normalize the so-called **free-photon number** $\tilde{\mathcal{H}}_0 / \hbar\omega$, i.e.,

$$\begin{aligned}
\text{normalized free-photon number} &\overset{\text{def}}{=\!=} \frac{\text{free-photon number}}{\left\langle \psi^{\mathrm{TV}} \middle| \psi^{\mathrm{TV}} \right\rangle} \\
&=\!= \frac{\tilde{\mathcal{H}}_0 \big/ \hbar\omega}{\left\langle \psi^{\mathrm{TV}} \middle| \psi^{\mathrm{TV}} \right\rangle},
\end{aligned} \tag{33}$$

called **normalized free-photon number**. Here, we emphasize that: the reason to call the photons "free" is that the $\tilde{\mathcal{H}}_0$ is just the radiation energy of the electromagnetic system [4]. In fact, we also have that

$$\begin{aligned}
\frac{1}{(33)} &= \frac{\left\langle \psi^{\mathrm{TV}} \middle| \hbar\omega \middle| \psi^{\mathrm{TV}} \right\rangle}{\left\langle \psi^{\mathrm{TV}} \middle| \tilde{\mathrm{H}}_0^{\mathrm{TV}} \middle| \psi^{\mathrm{TV}} \right\rangle} \\
&= \frac{\left\langle \psi^{\mathrm{TV}} \middle\| \Lambda_{\pm\hbar\omega}^{\mathrm{TV}} \middle\| \psi^{\mathrm{TV}} \right\rangle}{\left\langle \psi^{\mathrm{TV}} \middle| \tilde{\mathrm{H}}_0^{\mathrm{TV}} \middle| \psi^{\mathrm{TV}} \right\rangle} \\
&= \frac{\left\langle \psi_{\mathrm{cap}}^{\mathrm{TV}} \middle| \Lambda_{\pm\hbar\omega}^{\mathrm{TV}} \middle| \psi_{\mathrm{cap}}^{\mathrm{TV}} \right\rangle - \left\langle \psi_{\mathrm{ind}}^{\mathrm{TV}} \middle| \Lambda_{\pm\hbar\omega}^{\mathrm{TV}} \middle| \psi_{\mathrm{ind}}^{\mathrm{TV}} \right\rangle}{\left\langle \psi^{\mathrm{TV}} \middle| \tilde{\mathrm{H}}_0^{\mathrm{TV}} \middle| \psi^{\mathrm{TV}} \right\rangle} \\
&= \frac{\left\langle \psi_{\mathrm{cap}} \middle| \mathrm{L}_0 \middle| \psi_{\mathrm{cap}} \right\rangle - \left\langle \psi_{\mathrm{ind}} \middle| \mathrm{L}_0 \middle| \psi_{\mathrm{ind}} \right\rangle}{\left\langle \psi \middle| \tilde{\mathrm{H}}_0 \middle| \psi \right\rangle} \\
&= \Theta.
\end{aligned} \tag{34}$$

Here, the first equality is based on (30); the second equality is based on (27); the third equality is based on (21), (30), eigenmodal orthogonality, and the **important conclusions** presented at the end of Section IV-B; the fourth equality is based on (30); the fifth equality is based on (17) and (22).

Thus, we immediately obtain the following valuable conclusions:

$$\boxed{\frac{1}{\Theta} = \text{normalized free-photon number} = \frac{\text{free-photon number}}{\left\langle \psi^{\mathrm{TV}} \middle| \psi^{\mathrm{TV}} \right\rangle}}, \tag{35}$$

i.e.,

1) the definition for the generalized quality factor Θ in (22) [9, Eq. (14)], [10, Eq. (18)] has a solid physical and mathematical foundation;

2) the physical essence of the generalized quality factor $\Theta$ is the reciprocal of normalized free-photon number.

In fact, these conclusions are the natural generalizations for the $\Theta$-related conclusions given in [9] and [10].

### D. Alternative Calculation Formulation for $\Theta$

If we use (22) to calculate the $\Theta$, we need to do the MMQ and the MMQ-based modal three-component eigen-decomposition beforehand. Now, we propose another more direct calculation formulation for $\Theta$ as follows:

$$
\begin{aligned}
\Theta &= \hbar\omega \cdot \frac{\overbrace{\left\langle \psi^{\mathrm{T}} \right| \left(\mathrm{V}^{-1}\right)^{\dagger}}^{\left\langle \psi^{\mathrm{TV}} \right|} \cdot \overbrace{\mathrm{V}^{-1} \left| \psi^{\mathrm{T}} \right\rangle}^{\left| \psi^{\mathrm{TV}} \right\rangle}}{\left\langle \psi \right| \tilde{\mathrm{H}}_0 \left| \psi \right\rangle} \\
&= \hbar\omega \cdot \frac{\left\langle \psi^{\mathrm{T}} \right| \left[ \left(\hbar\omega\right)^{-\frac{1}{2}} \left|\Lambda^{\mathrm{TU}}\right|^{\frac{1}{2}} \cdot \mathrm{U}^{-1} \right]^{\dagger} \cdot \left(\hbar\omega\right)^{-\frac{1}{2}} \left|\Lambda^{\mathrm{TU}}\right|^{\frac{1}{2}} \cdot \mathrm{U}^{-1} \left| \psi^{\mathrm{T}} \right\rangle}{\left\langle \psi \right| \tilde{\mathrm{H}}_0 \left| \psi \right\rangle} \\
&= \frac{\left\langle \psi^{\mathrm{T}} \right| \mathrm{U} \cdot \left|\Lambda^{\mathrm{TU}}\right| \cdot \mathrm{U}^{\dagger} \left| \psi^{\mathrm{T}} \right\rangle}{\left\langle \psi \right| \tilde{\mathrm{H}}_0 \left| \psi \right\rangle} \\
&= \frac{\left\langle \psi \right| \overbrace{\left(\mathrm{T}^{-1}\right)^{\dagger} \cdot \mathrm{U}}^{\left[(\mathrm{T}\cdot\mathrm{U})^{-1}\right]^{\dagger}} \cdot \left|\Lambda^{\mathrm{TU}}\right| \cdot \overbrace{\mathrm{U}^{\dagger} \cdot \mathrm{T}^{-1}}^{(\mathrm{T}\cdot\mathrm{U})^{-1}} \left| \psi \right\rangle}{\left\langle \psi \right| \tilde{\mathrm{H}}_0 \left| \psi \right\rangle} \\
&= \boxed{\frac{\left\langle \psi \right| \tilde{\mathrm{L}}_0 \left| \psi \right\rangle}{\left\langle \psi \right| \tilde{\mathrm{H}}_0 \left| \psi \right\rangle}}.
\end{aligned} \tag{36}
$$

Here, the first equality is based on (29) and (34); the second equality is based on (26); the third equality is based on $\mathrm{U}^{-1} = \mathrm{U}^{\dagger}$, because the $\mathrm{U}$ is an unitary matrix; the fourth equality is based on (23); the last equality is based on the following definition:

$$
\boxed{\tilde{\mathrm{L}}_0 \overset{\mathrm{def}}{=\!=} \left[\left(\mathrm{T}\cdot\mathrm{U}\right)^{-1}\right]^{\dagger} \cdot \left|\Lambda^{\mathrm{TU}}\right| \cdot \left(\mathrm{T}\cdot\mathrm{U}\right)^{-1}}, \tag{37}
$$

which is called **matrix-formed generalized Lagrangian** in this paper, and it will be more carefully discussed in Section V-A. Obviously, the (36) does not need to do the MMQ and MMQ-based modal three-component eigen-decomposition before calculating the generalized quality factor $\Theta$.

Obviously, when the electromagnetic system works at an eigen-state $|\psi_{\xi}>$ (i.e., $|\psi>=|\psi_{\xi}>$), the generalized quality factor $\Theta$ is specified as follows:

$$
\Theta \xrightarrow{\left|\psi\right\rangle = \left|\psi_{\xi}\right\rangle} \left|\theta_{\xi}\right| \xrightarrow{\text{denoted by}} \vartheta_{\xi} . \tag{38}
$$

Thus, the eigenmodal quanta can also be obtained from the following eigen-equations:

$$
\boxed{\tilde{\mathrm{L}}_0 \left|\psi_{\xi}\right\rangle = \vartheta_{\xi} \cdot \tilde{\mathrm{H}}_0 \left|\psi_{\xi}\right\rangle}. \tag{39}
$$

The eigenmodal quanta $|\psi_{\xi}>$ calculated from the eigen-equation (39) are the same as the ones calculated from the eigen-equations (3) and (16).

## V. Discussions

In this section, we will discuss some important topics closely related to the MMQ and generalized quality factor $\Theta$.

### A. Generalized Lagrange–Hamilton Relation

From (36), it is immediate to obtain the following

$$
\boxed{\tilde{\mathcal{L}}_0 = \Theta \cdot \tilde{\mathcal{H}}_0}, \tag{40}
$$

where

$$
\left.
\begin{aligned}
\tilde{\mathcal{L}}_0 &= \left\langle \psi \right| \tilde{\mathrm{L}}_0 \left| \psi \right\rangle, \\
\tilde{\mathcal{H}}_0 &= \left\langle \psi \right| \tilde{\mathrm{H}}_0 \left| \psi \right\rangle.
\end{aligned}
\right\} \tag{41}
$$

This paper calls the $\tilde{\mathcal{L}}_0$ **generalized Lagrangian** to distinguish it from the classical Lagrangian $\mathcal{L}_0$ (the eigen-values of $\tilde{\mathrm{L}}_0$ are the absolute values of the eigen-values of $\mathrm{L}_0$).

Now, we simply compare the classical Lagrangian $\mathcal{L}_0$ and the generalized Lagrangian $\tilde{\mathcal{L}}_0$ as follows:

$$
\left.
\begin{aligned}
\Theta &\neq \frac{\mathcal{L}_0}{\tilde{\mathcal{H}}_0} = \frac{\left\langle \psi \right| \mathrm{L}_0 \left| \psi \right\rangle}{\left\langle \psi \right| \tilde{\mathrm{H}}_0 \left| \psi \right\rangle} \xrightarrow{\left|\psi\right\rangle = \left|\psi_{\xi}\right\rangle} \frac{\mathcal{L}_{0;\xi}}{\tilde{\mathcal{H}}_{0;\xi}} = \theta_{\xi} = \begin{cases} +\vartheta_{\xi}, \text{ if } \theta_{\xi} > 0, \\ \pm\vartheta_{\xi}, \text{ if } \theta_{\xi} = 0, \\ -\vartheta_{\xi}, \text{ if } \theta_{\xi} < 0; \end{cases} \\
\Theta &= \frac{\tilde{\mathcal{L}}_0}{\tilde{\mathcal{H}}_0} = \frac{\left\langle \psi \right| \tilde{\mathrm{L}}_0 \left| \psi \right\rangle}{\left\langle \psi \right| \tilde{\mathrm{H}}_0 \left| \psi \right\rangle} \xrightarrow{\left|\psi\right\rangle = \left|\psi_{\xi}\right\rangle} \frac{\tilde{\mathcal{L}}_{0;\xi}}{\tilde{\mathcal{H}}_{0;\xi}} = \vartheta_{\xi} = \left|\theta_{\xi}\right| \geq 0.
\end{aligned}
\right\} \tag{42}
$$

In addition, from the conclusions given in Section IV we have

$$
\frac{\mathcal{L}_0}{\tilde{\mathcal{H}}_0} \leq \frac{\left|\mathcal{L}_0\right|}{\tilde{\mathcal{H}}_0} \leq \frac{\tilde{\mathcal{L}}_0}{\tilde{\mathcal{H}}_0} = \Theta . \tag{43}
$$

Here, in the first inequality, the equality holds if and only if $|\psi_{\mathrm{ind}}>=0$; in the second inequality, the equality holds if and only if $|\psi_{\mathrm{ind}}>=0$ or $|\psi_{\mathrm{cap}}>=0$.

### B. Physical Essence of the Eigen-Value $\theta_{\xi}$

Employing (3), (16), (17), and (42), it is easy to derive the following relationships:

$$
\theta_{\xi} = \left\{
\begin{aligned}
&\frac{\left\langle \psi_{\xi} \right| \mathrm{P}_0^{-} \left| \psi_{\xi} \right\rangle}{\left\langle \psi_{\xi} \right| \mathrm{P}_0^{+} \left| \psi_{\xi} \right\rangle} = \frac{\mathrm{Im}\left\{\mathcal{P}_{0;\xi}\right\}}{\mathrm{Re}\left\{\mathcal{P}_{0;\xi}\right\}}, \\
&\frac{\left\langle \psi_{\xi} \right| \mathrm{L}_0 \left| \psi_{\xi} \right\rangle}{\left\langle \psi_{\xi} \right| \tilde{\mathrm{H}}_0 \left| \psi_{\xi} \right\rangle} = \frac{\mathcal{L}_{0;\xi}}{\tilde{\mathcal{H}}_{0;\xi}}, \\
&\frac{1}{c} \cdot \frac{\left\langle \psi_{\xi} \right| \mathrm{L}_0 \left| \psi_{\xi} \right\rangle}{\left\langle \psi_{\xi} \right| \mathrm{G}_0 \left| \psi_{\xi} \right\rangle} = \frac{1}{c} \cdot \frac{\mathcal{L}_{0;\xi}}{\mathcal{G}_{0;\xi}}, \\
&\mathrm{sign}\left\{\theta_{\xi}\right\} \cdot \frac{\left\langle \psi_{\xi} \right| \tilde{\mathrm{L}}_0 \left| \psi_{\xi} \right\rangle}{\left\langle \psi_{\xi} \right| \tilde{\mathrm{H}}_0 \left| \psi_{\xi} \right\rangle} = \mathrm{sign}\left\{\theta_{\xi}\right\} \cdot \frac{\tilde{\mathcal{L}}_{0;\xi}}{\tilde{\mathcal{H}}_{0;\xi}}, \\
&\frac{\bar{\mathcal{S}}_{0;\xi}}{\hbar} = \frac{\mathcal{L}_{0;\xi}}{\hbar\omega} = \frac{\mathcal{L}_{0;\xi}}{\hbar k} \cdot \frac{1}{c} = \frac{\mathcal{L}_{0;\xi}}{pc} = \frac{\mathcal{L}_{0;\xi}}{mc^2}.
\end{aligned}
\right\} \tag{44}
$$

Here, the first line is the classical results; the other lines are the new results revealed in this series of papers; the $\mathrm{sign}\{\theta_\xi\}$ means the sign of $\theta_\xi$, and $\mathrm{sign}\{0\}=0$.

Based on the results and conclusions obtained in Section IV, it is immediate to obtain the following conclusion:

$$\begin{aligned}\frac{1}{\vartheta_\xi}&=\frac{1}{|\theta_\xi|}\\&=\frac{\text{free-photon number of the }\xi\text{th eigenmodal quantum}}{\langle\psi_\xi^{\mathrm{TV}}|\psi_\xi^{\mathrm{TV}}\rangle}\\&=\text{normalized free-photon number of the }\xi\text{th eigenmode.}\end{aligned}\tag{45}$$

This is just the physical essence of eigen-value $\theta_\xi$ revealed from the eigen-relations given in (44) and Sections II ~ IV.

In addition, if the $\omega_{\mathrm{res}}^{\mathrm{ext}}$ is an external resonance frequency of the eigenmodal quantum $|\psi_\xi>$, then we have that

$$\left.\begin{array}{r}\vartheta_\xi=|\theta_\xi|\\ \hbar\omega\cdot\langle\psi_\xi^{\mathrm{TV}}|\psi_\xi^{\mathrm{TV}}\rangle=\langle\psi_\xi^{\mathrm{TU}}\|\Lambda^{\mathrm{TU}}\|\psi_\xi^{\mathrm{TU}}\rangle\\ \vartheta_\xi^{-1}=|\theta_\xi^{-1}|\\ \vartheta_\xi^{-1}\cdot\langle\psi_\xi^{\mathrm{TV}}|\psi_\xi^{\mathrm{TV}}\rangle=|\theta_\xi^{-1}|\cdot\langle\psi_\xi^{\mathrm{TV}}|\psi_\xi^{\mathrm{TV}}\rangle\end{array}\right\}\xrightarrow{\omega\to\omega_{\mathrm{res}}^{\mathrm{ext}}}\left\{\begin{array}{c}0,\\0,\\+\infty,\\ \dfrac{\tilde{\mathcal{H}}_{0;\xi}}{\hbar\omega_{\mathrm{res}}^{\mathrm{ext}}};\end{array}\right\}\tag{46}$$

if the $\omega_{\mathrm{res}}^{\mathrm{int}}$ is an internal resonance frequency of the eigenmodal quantum $|\psi_\xi>$, then we have that

$$\left.\begin{array}{r}\vartheta_\xi=|\theta_\xi|\\ \hbar\omega\cdot\langle\psi_\xi^{\mathrm{TV}}|\psi_\xi^{\mathrm{TV}}\rangle=\langle\psi_\xi^{\mathrm{TU}}\|\Lambda^{\mathrm{TU}}\|\psi_\xi^{\mathrm{TU}}\rangle\\ \vartheta_\xi^{-1}=|\theta_\xi^{-1}|\\ \vartheta_\xi^{-1}\cdot\langle\psi_\xi^{\mathrm{TV}}|\psi_\xi^{\mathrm{TV}}\rangle=|\theta_\xi^{-1}|\cdot\langle\psi_\xi^{\mathrm{TV}}|\psi_\xi^{\mathrm{TV}}\rangle\end{array}\right\}\xrightarrow{\omega\to\omega_{\mathrm{res}}^{\mathrm{int}}}\left\{\begin{array}{c}+\infty,\\0,\\0,\\0.\end{array}\right\}\tag{47}$$

Thus, any internally resonant eigenmodal quantum does not radiate free photon to infinity $\mathrm{S}_\infty$.

## C. Natural Normalization for Eigenmodal Quanta

As shown in the previous (44), we have the following relationships:

$$\left\{\begin{array}{c}\mathcal{L}_{0;\xi}/\tilde{\mathcal{H}}_{0;\xi}\\ \mathcal{L}_{0;\xi}/(c\cdot\mathcal{G}_{0;\xi})\end{array}\right\}=\theta_\xi=\left\{\begin{array}{c}\mathcal{L}_{0;\xi}/\hbar\omega\\ \mathcal{L}_{0;\xi}/(c\cdot\hbar k)\end{array}\right\},\tag{48}$$

and then have the following relationships:

$$\left.\begin{array}{l}\tilde{\mathcal{H}}_{0;\xi}=\hbar\cdot\omega=mc^2,\\ \mathcal{G}_{0;\xi}=\hbar\cdot k=mc.\end{array}\right\}\tag{49}$$

This means that: a natural normalization for the eigenmodal quanta is to normalize the eigen-Hamiltonian $\tilde{\mathcal{H}}_{0;\xi}$ to the single-photon energy $\hbar\omega$, or equivalently to normalize the eigen-momentum $\mathcal{G}_{0;\xi}$ to the single-photon momentum $\hbar k$.

In addition, the previous (33), (34), and (49) immediately give the following relationships:

$$\begin{array}{ccccc}\Theta & \xrightarrow{|\psi\rangle=|\psi_\xi\rangle} & |\theta_\xi| & = & \vartheta_\xi\\ \| & & & & \|\\ \dfrac{\langle\psi^{\mathrm{TV}}|\psi^{\mathrm{TV}}\rangle}{\tilde{\mathcal{H}}_0/\hbar\omega} & \xrightarrow{|\psi\rangle=|\psi_\xi\rangle} & \dfrac{\langle\psi_\xi^{\mathrm{TV}}|\psi_\xi^{\mathrm{TV}}\rangle}{\tilde{\mathcal{H}}_{0;\xi}/\hbar\omega} & = & \langle\psi_\xi^{\mathrm{TV}}|\psi_\xi^{\mathrm{TV}}\rangle\end{array}\tag{50}$$

i.e., the $\tilde{\mathcal{H}}_{0;\xi}$-oriented natural normalization (49) is equivalent to the following $|\psi_\xi^{\mathrm{TV}}>$-oriented natural normalization:

$$\langle\psi_\xi^{\mathrm{TV}}|\psi_\xi^{\mathrm{TV}}\rangle=|\theta_\xi|=\vartheta_\xi,\tag{51}$$

and this is completely consistent with that: when $\tilde{\mathcal{H}}_{0;\xi}=\hbar\omega$, the free-photon number of the $\xi$th eigenmodal quantum is 1, but the normalized free-photon number $(\tilde{\mathcal{H}}_{0;\xi}/\hbar\omega)/<\psi_\xi^{\mathrm{TV}}|\psi_\xi^{\mathrm{TV}}>$ is $1/|\theta_\xi|$ $(=1/\vartheta_\xi)$.

Here, it is necessary to emphasize that: some relationships in (48)–(51) are not applicable to the internally resonant eigenmodal quanta, because the eigen-Hamiltonians of the internally resonant eigenmodal quanta are 0, and then they cannot be normalized to $\hbar\omega$.

## D. Generalized Parseval Identity

In [9, Eq. (12)] and [10, Eq. (4)], an MMQ-oriented **Parseval identity** was derived. In this subsection, the Parseval identity is further extended to a generalized version.

First, any stationary-state mode $|\psi>$ can be linearly expanded in terms of the eigenmodal quanta $|\psi_\xi>$ as follows:

$$|\psi\rangle=\sum_\xi c_\xi|\psi_\xi\rangle,\tag{52}$$

where the $c_\xi$ is the expansion coefficient corresponding to the $\xi$th eigenmodal quantum.

Based on the discussions given in Section IV, it is easy to calculate the free-photon number carried by any mode $|\psi>$ as follows:

$$\begin{aligned}\text{free-photon number}&=\!=\!=\frac{1}{\hbar\omega}\cdot\overbrace{\langle\psi|\tilde{\mathrm{H}}_0|\psi\rangle}^{\tilde{\mathcal{H}}_0}\\&=\!=\!=\frac{1}{\hbar\omega}\cdot\sum_\xi|c_\xi|^2\,\tilde{\mathcal{H}}_{0;\xi}\\&\overset{(49)}{=\!=\!=}\sum_\xi|c_\xi|^2,\end{aligned}\tag{53}$$

called **generalized Parseval identity**. Here, the first equality is based on Section IV-C; the second and third equalities are based on (52), the eigenmodal orthogonality, and the natural normalization (49).

## E. Lower Bound of Generalized Quality Factor

In this sub-section, we will prove that: the lower bound of generalized quality factor Θ is the same as the smallest value of $|\theta_\xi|$ $(=\vartheta_\xi)$, i.e., $\min\{\Theta\}=\min\{|\theta_\xi|\}=\min\{\vartheta_\xi\}$.

We order all of the $\vartheta_\xi$ $(=|\theta_\xi|)$ by magnitude from the smallest to the largest, as follows:

$$\overbrace{\underbrace{\tilde{\mathcal{L}}_{0;1}/\tilde{\mathcal{H}}_{0;1}}_{\vartheta_1}}^{\Phi_{|\psi_1\rangle}} < \overbrace{\underbrace{\tilde{\mathcal{L}}_{0;2}/\tilde{\mathcal{H}}_{0;2}}_{\vartheta_2}}^{\Phi_{|\psi_2\rangle}} < \overbrace{\underbrace{\tilde{\mathcal{L}}_{0;3}/\tilde{\mathcal{H}}_{0;3}}_{\vartheta_3}}^{\Phi_{|\psi_3\rangle}} < \cdots, \tag{54}$$

where $\tilde{\mathcal{L}}_{0;1} = |\mathcal{L}_{0;1}|$, $\tilde{\mathcal{L}}_{0;2} = |\mathcal{L}_{0;2}|$, and $\tilde{\mathcal{L}}_{0;3} = |\mathcal{L}_{0;3}|$. Thus, we have the following relationships:

$$\left.\begin{aligned} 0 &< \frac{\tilde{\mathcal{H}}_{0;2}}{\tilde{\mathcal{H}}_{0;1}} < \frac{\tilde{\mathcal{L}}_{0;2}}{\tilde{\mathcal{L}}_{0;1}}, \\ 0 &< \frac{\tilde{\mathcal{L}}_{0;1}}{\tilde{\mathcal{L}}_{0;2}} < \frac{\tilde{\mathcal{H}}_{0;1}}{\tilde{\mathcal{H}}_{0;2}}, \end{aligned}\right\} \tag{55}$$

where we suppose that the $|\psi_1>$ and $|\psi_2>$ are neither internally resonant not externally resonant.

If we set $|\psi>=|c_1 \cdot \psi_1 + c_2 \cdot \psi_2>$, the generalized quality factor $\Theta_{|c_1\cdot\psi_1+c_2\cdot\psi_2>}$ satisfies the following relationships:

$$\begin{aligned} \vartheta_1 &= \tilde{\mathcal{L}}_{0;1}/\tilde{\mathcal{H}}_{0;1} \\ &= \frac{\tilde{\mathcal{L}}_{0;1}}{\tilde{\mathcal{H}}_{0;1}} \cdot \frac{|c_1|^2 + |c_2|^2 \cdot \tilde{\mathcal{L}}_{0;2}/\tilde{\mathcal{L}}_{0;1}}{|c_1|^2 + |c_2|^2 \cdot \tilde{\mathcal{L}}_{0;2}/\tilde{\mathcal{L}}_{0;1}} \\ &< \frac{\tilde{\mathcal{L}}_{0;1}}{\tilde{\mathcal{H}}_{0;1}} \cdot \frac{|c_1|^2 + |c_2|^2 \cdot \tilde{\mathcal{L}}_{0;2}/\tilde{\mathcal{L}}_{0;1}}{|c_1|^2 + |c_2|^2 \cdot \tilde{\mathcal{H}}_{0;2}/\tilde{\mathcal{H}}_{0;1}} \\ &= \frac{|c_1|^2 \cdot \tilde{\mathcal{L}}_{0;1} + |c_2|^2 \cdot \tilde{\mathcal{L}}_{0;2}}{|c_1|^2 \cdot \tilde{\mathcal{H}}_{0;1} + |c_2|^2 \cdot \tilde{\mathcal{H}}_{0;2}} \\ &= \Theta_{|c_1\cdot\psi_1+c_2\cdot\psi_2\rangle}. \end{aligned} \tag{56}$$

Here, the first two equalities are obvious; the third equality is based on (55); the last equality is based on the definition (22). Similarly, we also have that

$$\begin{aligned} \Theta_{|c_1\cdot\psi_1+c_2\cdot\psi_2\rangle} &= \frac{|c_1|^2 \cdot \tilde{\mathcal{L}}_{0;1} + |c_2|^2 \cdot \tilde{\mathcal{L}}_{0;2}}{|c_1|^2 \cdot \tilde{\mathcal{H}}_{0;1} + |c_2|^2 \cdot \tilde{\mathcal{H}}_{0;2}} \\ &= \frac{\tilde{\mathcal{L}}_{0;2}}{\tilde{\mathcal{H}}_{0;2}} \cdot \frac{|c_1|^2 \cdot \tilde{\mathcal{L}}_{0;1}/\tilde{\mathcal{L}}_{0;2} + |c_2|^2}{|c_1|^2 \cdot \tilde{\mathcal{H}}_{0;1}/\tilde{\mathcal{H}}_{0;2} + |c_2|^2} \\ &< \frac{\tilde{\mathcal{L}}_{0;2}}{\tilde{\mathcal{H}}_{0;2}} \cdot \frac{|c_1|^2 \cdot \tilde{\mathcal{L}}_{0;1}/\tilde{\mathcal{L}}_{0;2} + |c_2|^2}{|c_1|^2 \cdot \tilde{\mathcal{L}}_{0;1}/\tilde{\mathcal{L}}_{0;2} + |c_2|^2} \\ &= \tilde{\mathcal{L}}_{0;2}/\tilde{\mathcal{H}}_{0;2} \\ &= \vartheta_2, \end{aligned} \tag{57}$$

whose proof is similar to the proof for (56).

From the above (56) and (57), it is immediate to obtain the following inequality relationships:

$$\min\{\vartheta_1, \vartheta_2\} = \vartheta_1 < \Theta_{|c_1\cdot\psi_1+c_2\cdot\psi_2\rangle} < \vartheta_2 = \max\{\vartheta_1, \vartheta_2\}. \tag{58}$$

This means that the $\Theta_{|c_1\cdot\psi_1+c_2\cdot\psi_2>}$ is larger than the smallest one and smaller than the largest one in $\{\vartheta_1, \vartheta_2\}$.

Similarly to proving the above (58), we can also prove the following more general inequality relationships:

$$\vartheta_{\xi_1} < \Theta_{|c_{\xi_1}\cdot\psi_{\xi_1}+c_{\xi_2}\cdot\psi_{\xi_2}+\cdots+c_{\xi_N}\cdot\psi_{\xi_N}\rangle} < \vartheta_{\xi_N}, \tag{59}$$

for any $\vartheta_{\xi_1} < \vartheta_{\xi_2} < \cdots < \vartheta_{\xi_N}$.

Thus, we obtain the conclusion that: the lower bound of generalized quality factor $\Theta$ is $\min\{|\theta_\xi|\}$, i.e.,

$$\min\{\Theta\} = \min\{\vartheta_\xi\} = \vartheta_1 = |\theta_1| = \min\{|\theta_\xi|\}. \tag{60}$$

This is precisely one of the core values for seeking resonant eigenmodal quanta via MMQ.

## VI. Numerical Experiments

In this section, we do the MMQ for the perfect electric conducting (PEC) parabolic reflector antenna reported in [26]. The radius of the reflector antenna is 31 cm, and the focal depth of the reflector antenna is 11.5 cm. The feeding port for the reflector antenna is a circular port with a radius of 2 cm, and the center of the port is placed in the focal point.

Here, we use the classical eigen-equation (16) [10, Eq. (14)] and the novel eigen-equation (39) to calculate the eigenmodal quanta of the reflector antenna. The modal significances (MSs) $1/|1+\mathrm{i}\cdot\theta_\xi|$ ($=1/|1+\mathrm{i}\cdot\vartheta_\xi|$) and absolute eigen-values $|\theta_\xi|$ ($=\vartheta_\xi$) of the first two eigenmodal quanta $|\psi_1>$ and $|\psi_2>$ are shown in Figs. 1 and 2, respectively. Obviously, the results calculated from the two different methods [10, Eq. (14)] and (39) are consistent with each other.

In addition, we also use the (22)-based classical formulation and the (36)-based novel formulation to calculate the generalized quality factors $\Theta$s of the modes $|2\cdot\psi_1+1\cdot\psi_2>$, $|1\cdot\psi_1+1\cdot\psi_2>$, and $|1\cdot\psi_1+2\cdot\psi_2>$, where the $|\psi_1>$ and $|\psi_2>$ are just the ones shown in Figs. 1 and 2. The $\Theta_{|2\cdot\psi_1+1\cdot\psi_2>}$, $\Theta_{|1\cdot\psi_1+1\cdot\psi_2>}$, and $\Theta_{|1\cdot\psi_1+2\cdot\psi_2>}$ calculated from the (22)-based and (36)-based formulations are shown in Figs. 3, 4, and 5, respectively. Obviously, the results calculated from (22) and (36) are consistent with each other, and, at the same time, the calculated $\Theta$s satisfy the inequality (58).

From the above numerical results, it is easy to observe that: the formulations (39) and (36) proposed in this paper are valid in doing MMQ and calculating generalized quality factor $\Theta$.

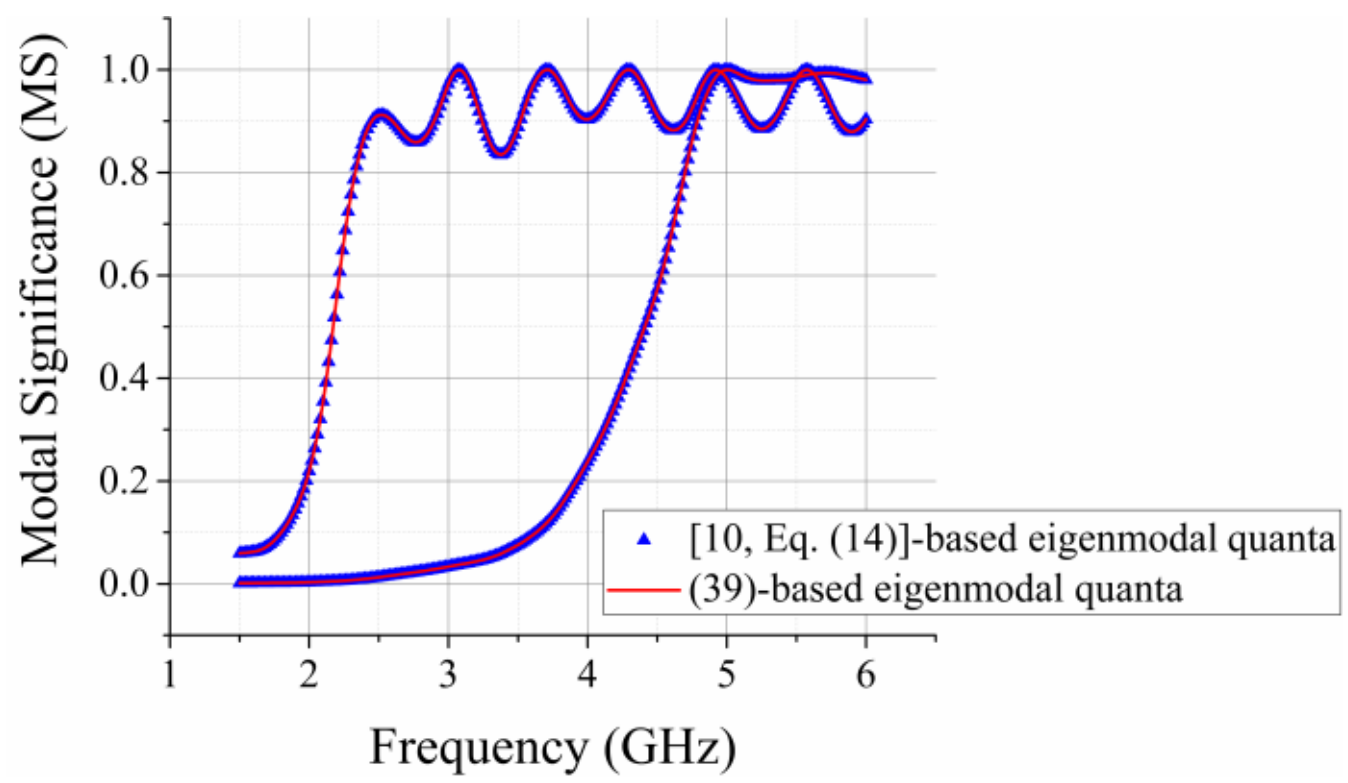

Fig. 1. [10, Eq. (14)]-based and (39)-based MSs of the first two eigenmodal quanta $|\psi_1>$ and $|\psi_2>$ of the reflector antenna reported in [26].

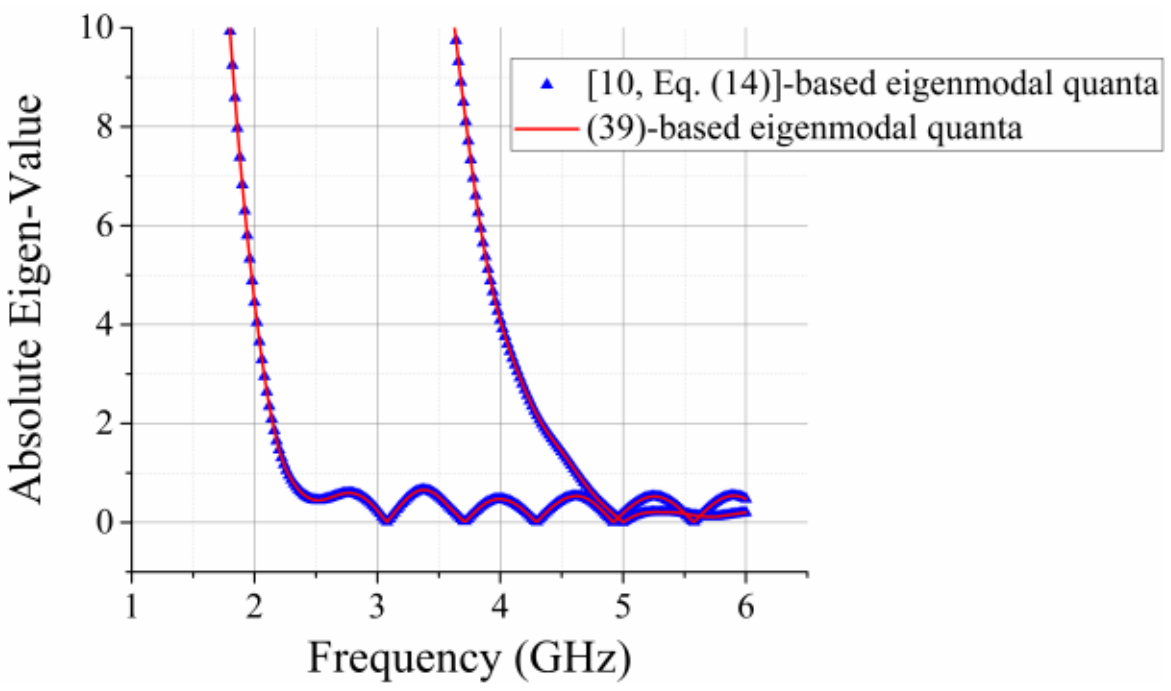

Fig. 2. [10, Eq. (14)]-based and (39)-based absolute eigen-values of the first two eigenmodal quanta $|\psi_1>$ and $|\psi_2>$ of the reflector antenna reported in [26].

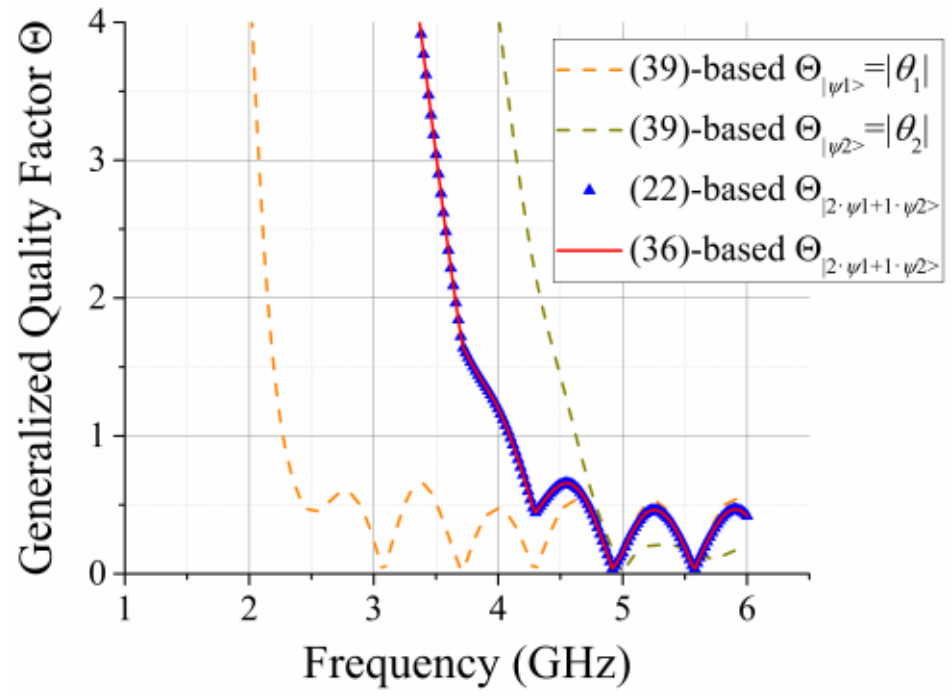

Fig. 3. (22)-based and (36)-based generalized quality factors $\Theta_{|2\cdot\psi_1+1\cdot\psi_2>}$ of $|2\cdot\psi_1+1\cdot\psi_2>$.

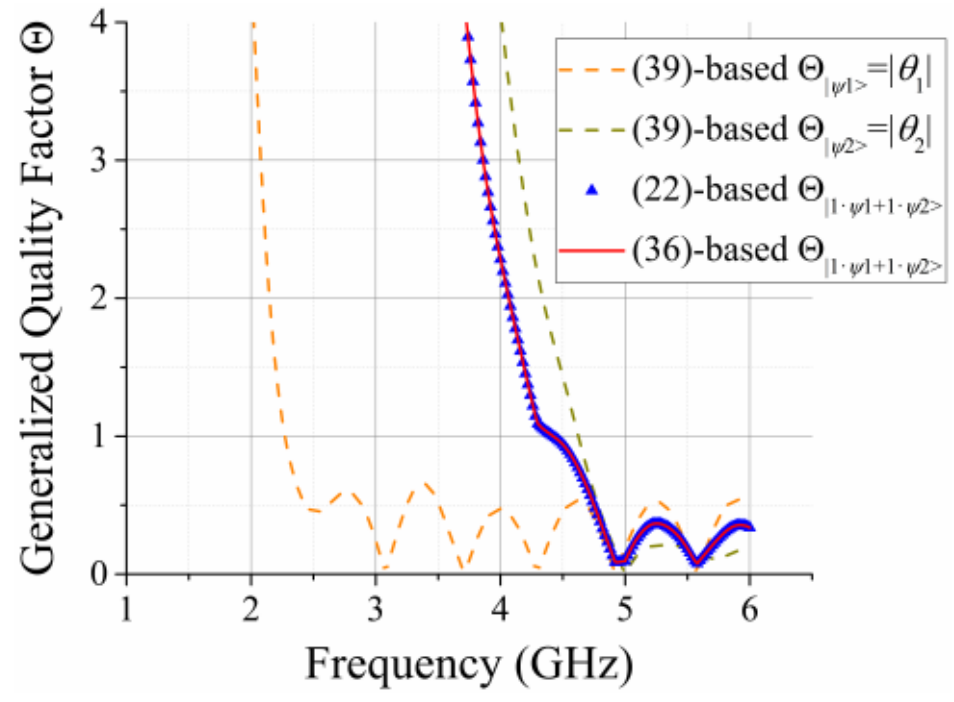

Fig. 4. (22)-based and (36)-based generalized quality factors $\Theta_{|1\cdot\psi_1+1\cdot\psi_2>}$ of $|1\cdot\psi_1+1\cdot\psi_2>$.

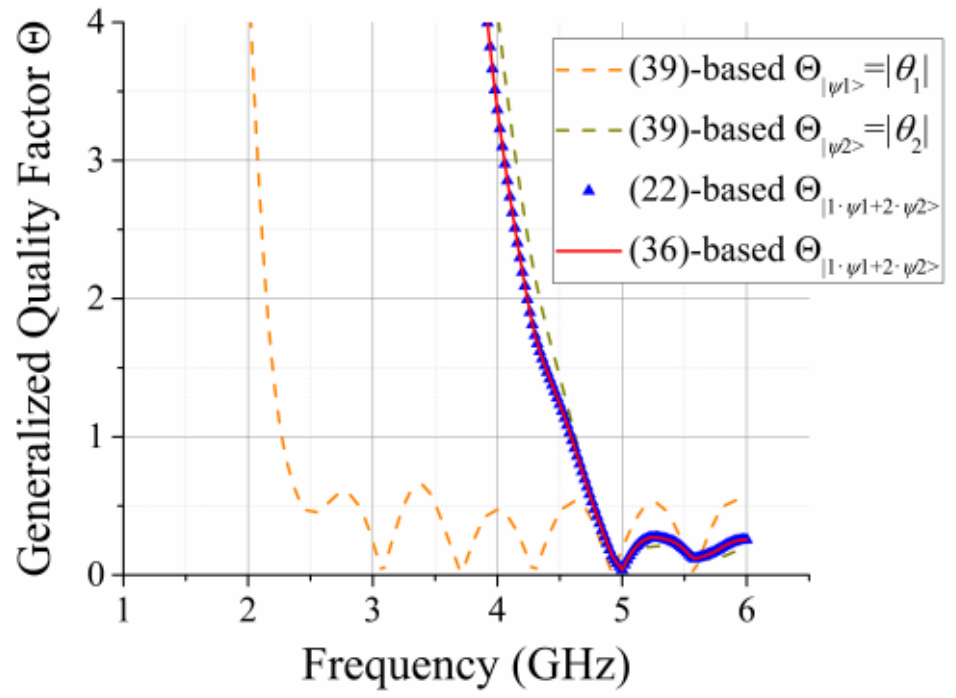

Fig. 5. (22)-based and (36)-based generalized quality factors $\Theta_{|1\cdot\psi_1+2\cdot\psi_2>}$ of $|1\cdot\psi_1+2\cdot\psi_2>$.

## VII. Summary and Conclusion

In our previous papers [4], we extended the classical Hamiltonian $\mathcal{H}$ to its generalized version—generalized Hamiltonian $\tilde{\mathcal{H}}_0$. In this paper, we further extend the classical Lagrangian $\mathcal{L}_0$ to its generalized version—generalized Lagrangian $\tilde{\mathcal{L}}_0$.

Employing the classical Lagrangian $\mathcal{L}_0$, generalized Lagrangian $\tilde{\mathcal{L}}_0$, and generalized Hamiltonian $\tilde{\mathcal{H}}_0$, this paper generalizes some well-known relations in microscopic world to their macroscopic versions:

1) we generalize the microscopic Planck–Einstein relation $E=\hbar\omega$ to its macroscopic version, and then obtain a so-called macroscopic action–energy eigen-relation $\omega\cdot\bar{\mathcal{S}}_0=\theta\cdot\tilde{\mathcal{H}}_0$, or alternatively called Lagrange–Hamilton eigen-relation $\mathcal{L}_0=\theta\cdot\tilde{\mathcal{H}}_0$;
2) we also extend the Lagrange–Hamilton eigen-relation $\mathcal{L}_0=\theta\cdot\tilde{\mathcal{H}}_0$ to a so-called generalized Lagrange–Hamilton relation $\tilde{\mathcal{L}}_0=\Theta\cdot\tilde{\mathcal{H}}_0$ (whose eigen version is $\tilde{\mathcal{L}}_0=\vartheta\cdot\tilde{\mathcal{H}}_0$);
3) we generalize the microscopic de Broglie relation $p=\hbar k$ to its macroscopic version, and then obtain macroscopic action–momentum eigen-relation $c^{-1}\cdot\mathcal{L}_0=\theta\cdot\mathcal{G}_0$;
4) in fact, the $c^{-1}\cdot\mathcal{L}_0=\theta\cdot\mathcal{G}_0$ can be further extended to the generalized action–momentum relation $c^{-1}\cdot\tilde{\mathcal{L}}_0=\Theta\cdot\mathcal{G}_0$;
5) we also reveal the equivalence between the above eigen-relations and the classical $\mathcal{P}_0^+-\mathcal{P}_0^-$ eigen-relation widely discussed in [5], [9], and [10].

These beautiful eigen-relations are summarized as follows:

$$\boxed{\begin{array}{ccccccccc} & & & & \mathcal{P}_0^+\cdot\cancel{\frac{\tau}{2}} & = & \frac{1}{\theta}\cdot\mathcal{P}_0^-\cdot\cancel{\frac{\tau}{2}}\cdot\frac{1}{\cancel{\omega}} & \cdot & \cancel{\omega} \\ & & & & \| & & \| & & \\ c & \cdot & \mathcal{G}_0 & = & \tilde{\mathcal{H}}_0 & = & \frac{1}{\theta}\cdot\mathcal{L}_0\cdot\frac{1}{\cancel{\omega}} & \cdot & \cancel{\omega} \\ & & \Updownarrow & & \Updownarrow & & \updownarrow & & \\ c & \cdot & p & = & E & = & \hbar & \cdot & \omega \\ & & \Updownarrow & & \Updownarrow & & \Updownarrow & & \\ c & \cdot & \mathcal{G}_0 & = & \tilde{\mathcal{H}}_0 & = & \frac{1}{\Theta}\cdot\tilde{\mathcal{L}}_0\cdot\frac{1}{\cancel{\omega}} & \cdot & \cancel{\omega} \end{array}} \tag{61}$$

In fact, the eigen-relations in (61) are just the **first principles** of macroscopic electrodynamics, because Maxwell equations can be derived from them as exhibited in [4].

Directly casting the eigen-relations into their operator forms yields three equivalent eigen-equations (16). Solving any one of the eigen-equations gives the eigenmodal quanta. Thus, the eigen-relations (61) and eigen-equations (16) become the first principles of MMQ, and then a first-principles-based **grand unified theory (GUT)** for MMQ is established. The first-principles-based GUT for MMQ indicates that:

1) for any macroscopic structure, its $\mathcal{L}_0$, $\tilde{\mathcal{L}}_0$, and $\tilde{\mathcal{H}}_0$ must satisfy the eigen-relations $\mathcal{L}_0=\theta\cdot\tilde{\mathcal{H}}_0$ and $\tilde{\mathcal{L}}_0=\vartheta\cdot\tilde{\mathcal{H}}_0$;
2) the intrinsic nature of the structure determines all feasible values of $\{\theta_\xi,\vartheta_\xi\}$, i.e., the eigen-values of the structure;
3) one eigen-value $\{\theta_\xi,\vartheta_\xi\}$ corresponds to one (non-degenerate) or more (degenerate) eigenmodal quanta;
4) the eigenmodal quanta are energy-decoupled, and can be

calculated by the first-principles-based eigen-equations;

5) the eigenmodal quanta span the whole Maxwellian space, i.e., any Maxwellian mode can be expanded in terms of the eigenmodal quanta.

Thus, the above first principles can be viewed as a **generalized action principle** for governing macroscopic electrodynamics.

As the theoretical applications of the first-principles-based GUT, this paper does the following works:

1) establishing a solid physical and mathematical foundation for the generalized quality factor Θ defined in [9]and[10];
   a) in our previous works [9, Eq. (14)], [10, Eq. (18)], we used macroscopic eigenmodal quanta to realize the three-component eigen-decomposition for any mode, and used the eigen-decomposition to define the Θ;
   b) however, the first principles underlying the above Θ-oriented definition was not revealed in the previous works;
   c) starting from the Lagrange–Hamilton eigen-relation, this paper derives the formulation for calculating normalized free-photon number, and proves that it equals to the reciprocal of Θ, and then reveals the first principles that underpin the aforementioned definition for Θ;
2) starting from the aforementioned first principle, this paper also derives a new calculation formula for Θ. Compared with the classical formula based on eigen-decomposition, the new formula is more concise and efficient;
3) introducing the concept of generalized Lagrangian $\tilde{\mathcal{L}}_0$, which is positively definite (the case of there is not resonant quantum) or positively semi-definite;
4) revealing the physical essence of Θ—the reciprocal of normalized free-photon number;
5) revealing the physical essence of eigen-value $\{\theta_\xi, \vartheta_\xi\}$ from several different points of view, i.e.,
   a) $\theta_\xi$ is the ratio of the macroscopic eigen-action $\bar{\mathcal{S}}_{0;\xi}$ to the microscopic action $\hbar$,
   b) $\theta_\xi$ ($\vartheta_\xi$) is the ratio of the classical eigen-Lagrangian $\mathcal{L}_{0;\xi}$ (the generalized eigen-Lagrangian $\tilde{\mathcal{L}}_{0;\xi}$) to the generalized eigen-Hamiltonian $\tilde{\mathcal{H}}_{0;\xi}$,
   c) $\vartheta_\xi$ ($=|\theta_\xi|$) is the absolute value of the generalized eigen-quality factor $\Theta \to |\theta_\xi|$ (here $|\psi> \to |\psi_\xi>$),
   d) $\vartheta_\xi$ ($=|\theta_\xi|$) is the reciprocal of the normalized free-photon number of the $\xi$th eigenmodal quantum, and
   e) the other ones listed in (44);
6) introducing a very natural way to normalize eigenmodal quanta—to normalize the eigen-Hamiltonian to the energy carried by a single photon;
7) generalizing the classical Parseval identity/theorem;
8) deriving the lower bound of the generalized quality factor Θ, and the lower bound is $\min\{\vartheta_\xi\}$ ($=\min\{|\theta_\xi|\}$).

In fact, besides the above listed theoretical applications, there are also many engineering applications from the first-principles-based GUT for MMQ, and the related developments will be reported in our subsequent papers.